\documentclass[aps,prl,reprint,groupedaddress]{revtex4-1}

\usepackage{graphicx}
\usepackage{dcolumn}
\usepackage{bm}

\usepackage{txfonts}

\begin{document}

\title{Theoretical Analysis on Pseudo-Degenerate Zero-Energy Modes in 
Vacancy-Centered Hexagonal Armchair Nanographene
}

\author{Naoki MORISHITA}\email{morishi@artemis.mp.es.osaka-u.ac.jp} 
\author{Gagus Ketut SUNNARDIANTO} 
\author{Satoaki MIYAO} 
\author{Koichi KUSAKABE\thanks{kabe@mp.es.osaka-u.ac.jp}}
\affiliation{Graduate School of Engineering Science, Osaka University,\\
1-3 Machikaneyama-cho, Toyonaka, Osaka, 560-8531, Japan
} 

\begin{abstract}
Deriving mathematical expressions of two zero modes for a $\pi$-band tight-binding model, we identify a class of bipartite graphs having the same number of subgraph sites, where each graph represents one of the quasi-hexagonal nanographene molecule with a center vacancy (VANG). Indeed, in a VANG molecule, C$_{60}$H$_{24}$, showing stability in a density-functional simulation at the highest occupied level, there appear two pseudo-degenerate zero modes, a vacancy-centered quasi-localized zero mode, and extending zero mode with a $\sqrt{3} \times \sqrt{3}$ structure. Since there is a finite energy gap between these two zero-energy modes and the other modes, low-lying states composed of quasi-degenerate zero modes appear as magnetic multiplets. Thus, the unique magnetic characteristics derived in our theory are expected to hold for synthesized VANG molecules in reality.
\end{abstract}

\date{\today}

\maketitle

\section{Introduction}
The physics and chemistry of topological zero modes found in graphene \cite{Novoselov2004, Novoselov2005, Geim2007, CastroNeto2009, Enoki2013, Aoki2014} are currently attracting a great deal of interest from researchers. One example of these zero modes is the edge state that appears at the zigzag edge (the Z$_1$ edge) of nanographene \cite{Fujita1996, Nakada1996}. The “topological” zero modes are usually localized around a specified Z$_1$ edge \cite{Klusek2000, Kobayashi2005, Niimi2005, Sugawara2006, Ziatdinov2013}, which may show interesting transport phenomena \cite{Wakabayashi2001, Wakabayashi2007} and correlation effects \cite{Wakabayashi1998, Kusakabe2003, Hikihara2003, Yoshioka2003, Son2006}. These edge states are now known to appear along various edges, including the Z$_1$ and A$_{21}$ edges \cite{Fujii2014, Ziatdinov2015}. 

Another category of these zero modes is defect-centered zero modes. Defect structures in the atomic scale, including bare vacancies (V) and triply-hydrogenated vacancies V$_{111}$, are examples of this\cite{Lehtinen2004, Pereira2006, Ziatdinov2014}. These also show quantum-mechanical magnetic signals\cite{Chen2011, Nair2012, McCreary2012}, another focus of research interest\cite{Yazyev2007, Ugeda2010, Haase2011, Uchoa2011, Sofo2012, Palacios2012, Kanao2012, Mitchell2013}.

There have been a lot of synthesis reports on polycyclic aromatic hydrocarbon molecules (PAHs)\cite{Clar1964, Wu2004} parallel to or prior to theoretical considerations\cite{Ezawa2007, Ezawa2008}. The precise fabrication of nanographene ribbon was achieved through a synthetic method\cite{Cai2010}. The movement has been further strengthened in the past few decades, motivated particularly by the discovery of graphene\cite{Novoselov2004, Novoselov2005, Geim2007}. Therefore, an approach for realizing these physical concepts as a chemical reality is required.

We found that two non-bonding molecular orbitals appear at zero-energy for a $\pi$ network system representing a polycyclic aromatic hydrocarbon. The Fermi level of the half-filled neutral system corresponds to zero-energy. The orbitals are a vacancy-centered quasi-localized zero mode (QLZM) and a $\sqrt{3} \times \sqrt{3}$ zero-energy extending mode (S3ZM), which are two-fold degenerate at $E=0$. This solution is defined for a series of molecular structures. With knowledge on the zero modes, we tested one of the molecular structures using density-functional-theory (DFT) simulations. The result demonstrates stability of the material. Since this smallest VANG structure is expected to be the most fragile among the series of VANG molecules, we conclude stability of the others. 

We also found that an energy eigenvalue problem for a $\pi$ electron system of a nanographene molecule is 
solvable in an algebraic way for the zero-energy case. Adopting a single-band tight-binding model (TBM) for the molecule, we show explicit construction of this exact solution. Once the topological nature of the non-bonding zero mode is fixed, we consider the ground state of the molecule. Using the known theorem on the Hubbard model \cite{Lieb1989} and the local-spin-density-approximation (LSDA) in DFT\cite{Hohenberg1964}, we conclude the singlet ground state for this molecule having quasi-degenerate highest-occupied molecular orbitals (QDHOMO). The topological nature of the zero modes suggests interesting magnetic properties in reality. 

\section{Vacancy-centered Hexagonal Armchair Nanographene}
\begin{figure}[tb]
\begin{tabular}{cc}
\begin{minipage}{0.5\hsize}
\begin{center}
\begin{flushleft}(a)\end{flushleft}
\includegraphics[width=4cm]{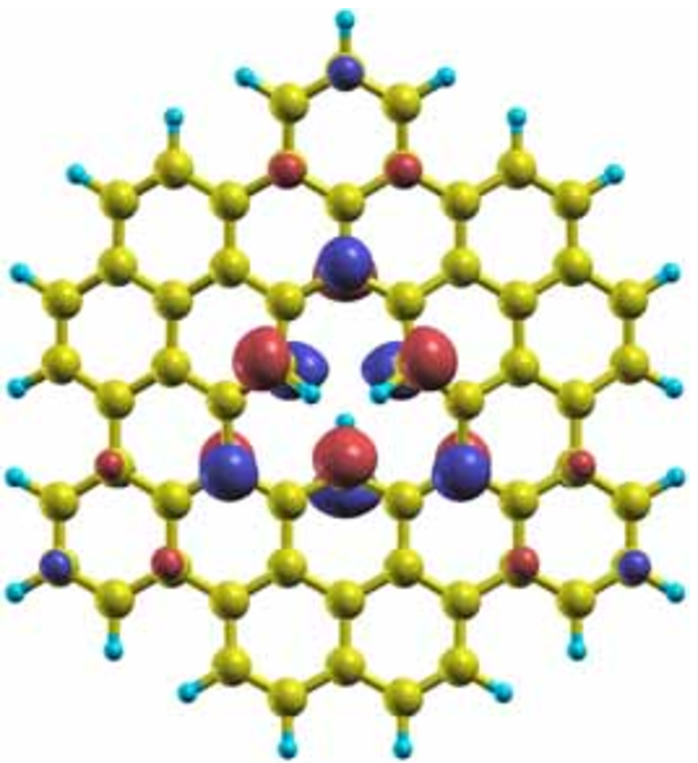}
\end{center}
\end{minipage}
\begin{minipage}{0.5\hsize}
\begin{center}
\begin{flushleft}(b)\end{flushleft}
\includegraphics[width=4cm]{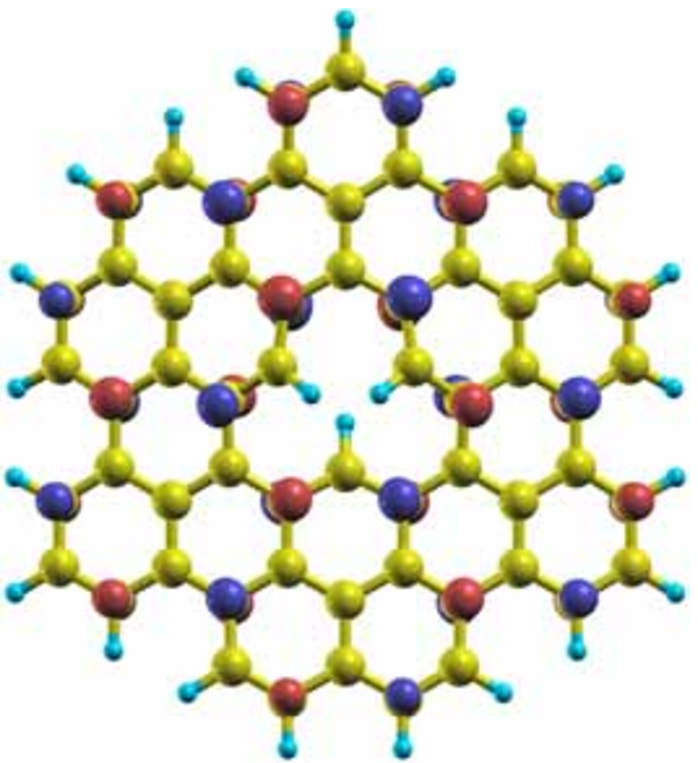}
\end{center} 
\end{minipage}
\end{tabular}
\caption{(Color) Vacancy-centered hexagonal armchair nanographene (VANG) molecule. This molecule has an armchair edge with $N_{\rm arm}=2$. (a) A  vacancy-centered quasi-localized zero mode (QLZM) is shown by iso surfaces (red surface for positive amplitudes and blue surfaces for negative amplitudes). (b) A $\sqrt{3} \times \sqrt{3}$ mode (S3ZM) is shown.}
\label{fig:DFT_VANG_NA2}
\end{figure}

A vacancy-centered hexagonal armchair nanographene (VANG) is a hexagonal nanographene molecular structure. It has a deformed hexagonal shape with an atomic vacancy at the center (Fig.~\ref{fig:DFT_VANG_NA2}). The structure contains a triply-hydrogenated vacancy known as V$_{111}$\cite{Ziatdinov2014}.  We introduce the quantity $N_{\rm arm}$, which is the number of armchair structures at an edge of the hexagon. The number of carbon atoms, $N_{\rm C}$ is given by  $N_{\rm C}=6N_{\rm arm}(3N_{\rm arm}-1)$. Note that the subgraph site numbers $N_{\rm A}$ and $N_{\rm B}$ satisfy $N_{\rm A}-N_{\rm B}=0$ for VANG. The armchair structure at the periphery consists of four carbon atoms, where each carbon is mono-hydrogenated. The size of the VANG molecule is denoted by$N_{\rm arm}$, which has to satisfy a condition $N_{\rm arm} \geq 2$. 

In Fig.~\ref{fig:DFT_VANG_NA2}, we show the two nearly degenerate orbitals at the Fermi level. The electronic structure was determined from a DFT simulation, the details of which are explained later. The first image is the topological QLZM around the vacancy (Fig.~\ref{fig:DFT_VANG_NA2}(a)). The other, Fig.~\ref{fig:DFT_VANG_NA2}(b), shows a $\sqrt{3} \times \sqrt{3}$ zero mode (S3ZM). The QLZM is an A$_1$ representation, while the S3ZM is an A$_2$ representation.

Adopting LDA\cite{Kohn1965, Perdew1992}, we performed structural optimization of a VANG molecule. The simulation is accurate enough for this particular purpose of the stability test. Super-cell simulation was undertaken using the plane-wave expansion method with an energy cutoff of 24 Ry for the wave function and the ultra-soft pseudo potential method\cite{Vanderbilt1990, Rappe1990}, in which a 240 Ry cutoff was used for the charge density expansion. The convergence criterion for the inter-atomic force ensures that the absolute value of the force vector in the multidimensional space becomes less than $1.0 \times 10^{-5}$ Ry/a.u. 

Similar to the V$_{111}$ structure of graphene, the VANG shows a deformed structure with reduced symmetry. A super-cell simulation using the {\sc Quantum ESPRESSO} code\cite{QE-2009} shows an optimized structure with $C_{1v}$ symmetry as displayed in Fig. \ref{fig:DFT_VANG_NA2} from XCrySDen\cite{Kokalj2003}. There appears two nearly degenerate zero modes, where QLZM appears below S3ZM by 26 meV. The molecular stability with the zero modes is clearly shown. Thus, the test ends with a positive result allowing a synthetic process to be considered.

\begin{figure}[b]
\begin{center}
\includegraphics[height=8.07cm]{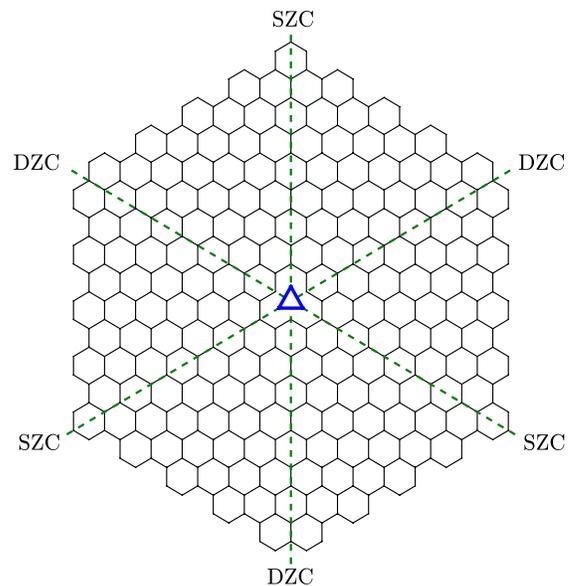}
\end{center}
\caption{(Color) A $\pi$-orbital network of a VANG molecule with $N_{\rm arm}=5$. This network has $C_{3v}$ point group symmetry. A triangle at the center represents a three-fold rotation axis and the three dashed lines show the mirror symmetry planes. Note that there are two types of corners, one is single-zigzag-corner (SZC), and the other is double-zigzag-corner (DZC).}
\label{fig:_C3vSymNA5}
\end{figure} 

\section{Analysis of QLZM}

The DFT result is not always convenient for the analysis of magnetic properties using an argument of the topology. We also adopt the tight-binding description of nanographene, which is often known as the single-orbital tight-binding model (TBM). This model is known to be valid for analysis of low-energy excitations of graphene and nanographene rather generally\cite{Enoki2013}. It is defined by a skeleton graph of the $\pi$ bond connections. The network of VANG has the structure outlined by Fig.~\ref{fig:_C3vSymNA5}. It has $C_{3v}$ point group symmetry.

\begin{figure}[t]
\begin{tabular}{cc}
\begin{minipage}{0.27\hsize}
\begin{center}
\begin{flushleft}(a)\end{flushleft} 
\includegraphics[height=4cm]{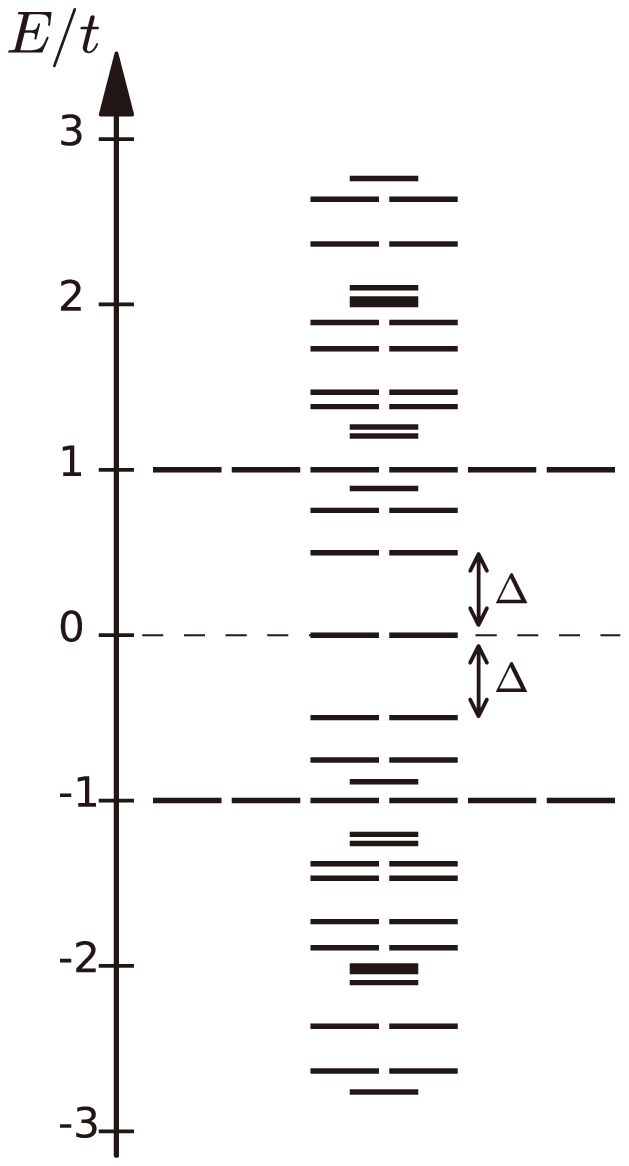}
\end{center}
\end{minipage}
\begin{minipage}{0.73\hsize}
\begin{center}
\begin{flushleft}(b)\end{flushleft}
\vspace{3mm}
\includegraphics[height=4cm]{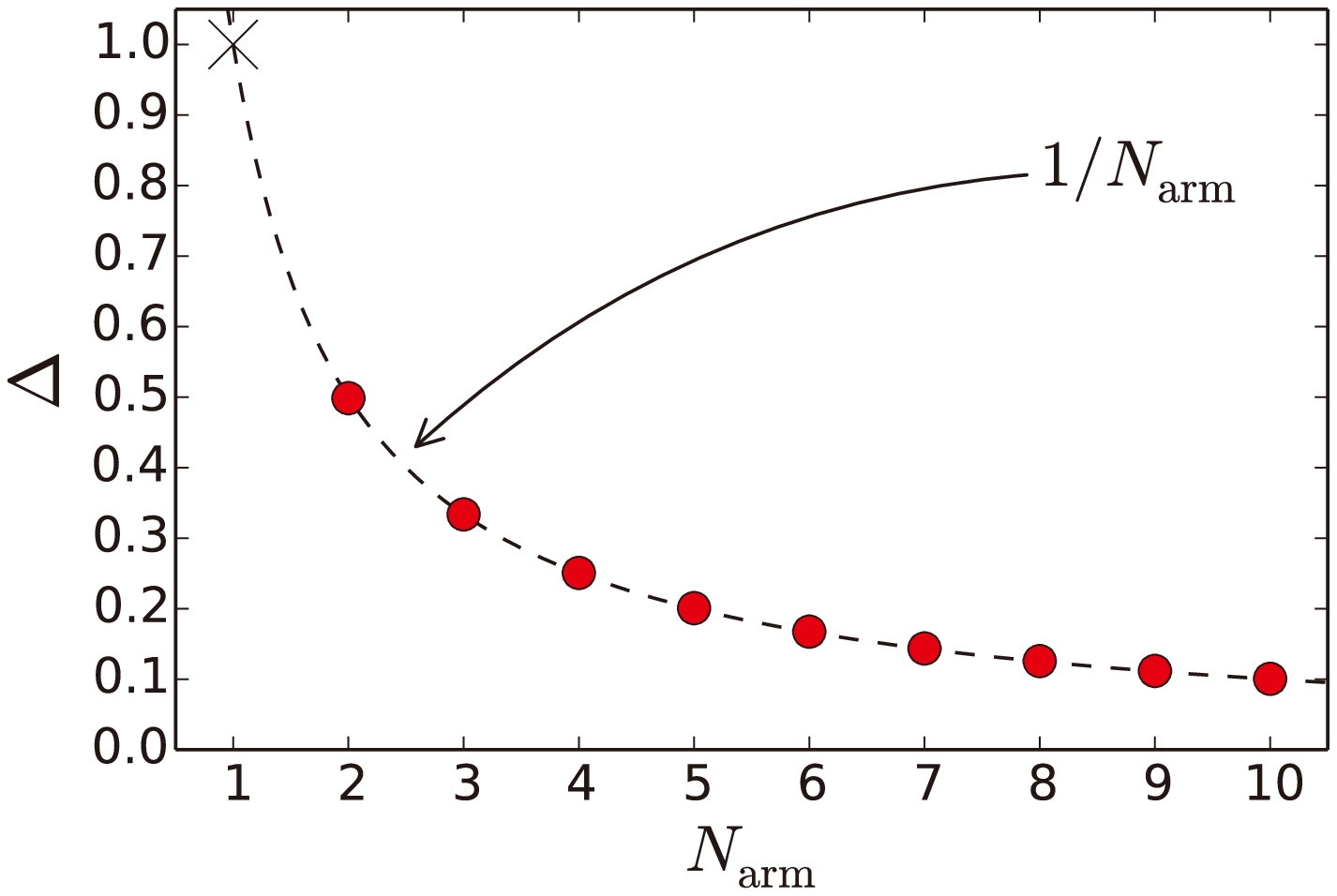}
\vspace{-3mm}
\end{center} 
\end{minipage}
\end{tabular}
\caption{(Color-online) (a) Eigen energies of the tight-binding model in case of $N_{\rm arm}=2 (N_{\rm C}=60)$. The energy at $E=0$ is doubly degenerate. Here, $t$($\sim$ 3 eV) is the nearest neighbor hopping parameter. (b) The size dependence of $\Delta$. The tight-binding result is very well fit by $1/N_{\rm arm}$. Note that $\Delta$ is renormalized by $t$.}
\label{fig:GapDelta}
\end{figure}

\begin{figure}[b]
\begin{tabular}{cc}
\begin{minipage}{0.5\hsize}
\begin{center}
\begin{flushleft}(a)\end{flushleft} 
\includegraphics[width=4cm]{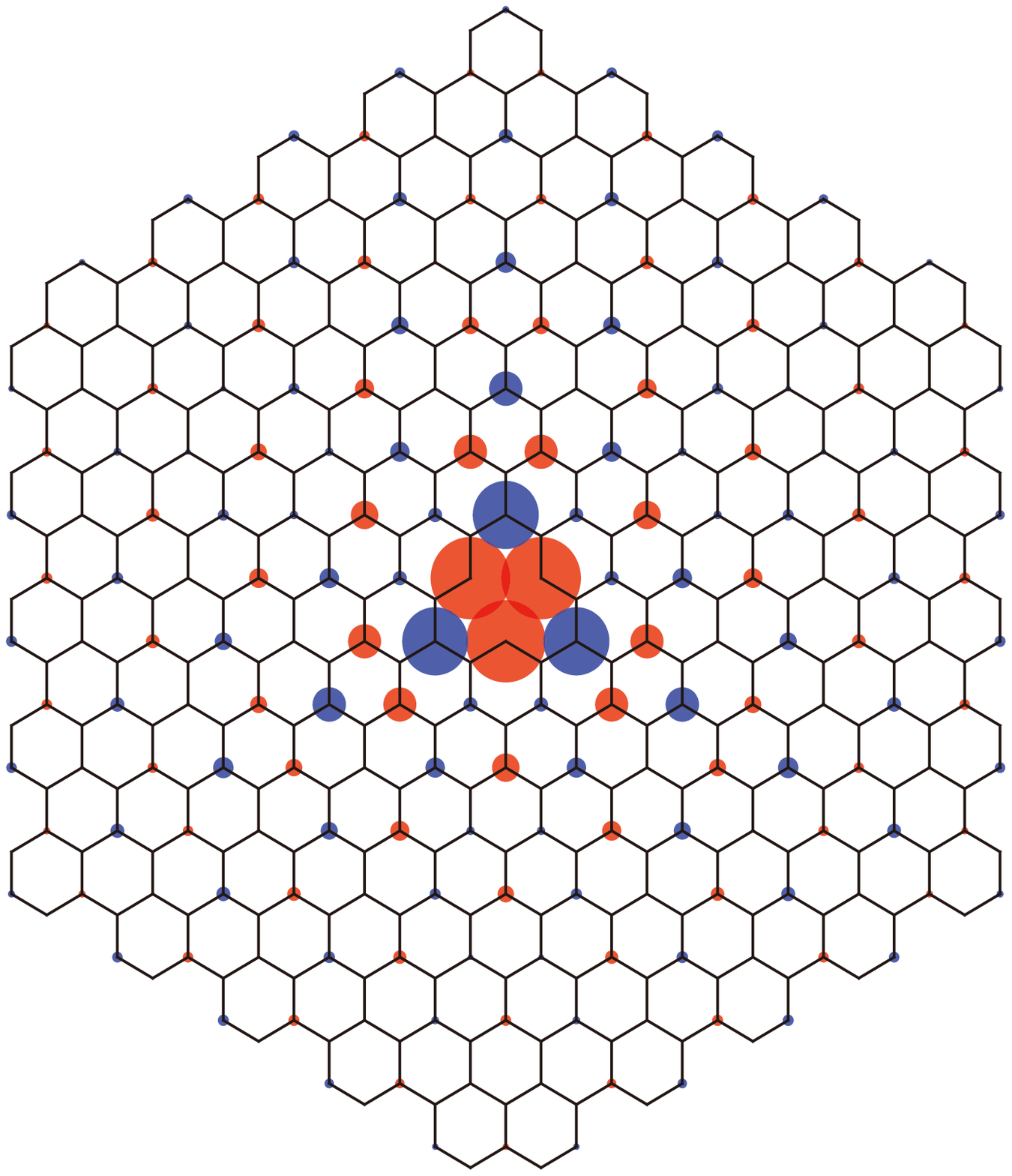}
\end{center}
\end{minipage}
\begin{minipage}{0.5\hsize}
\begin{center}
\begin{flushleft}(b)\end{flushleft}
\includegraphics[width=4cm]{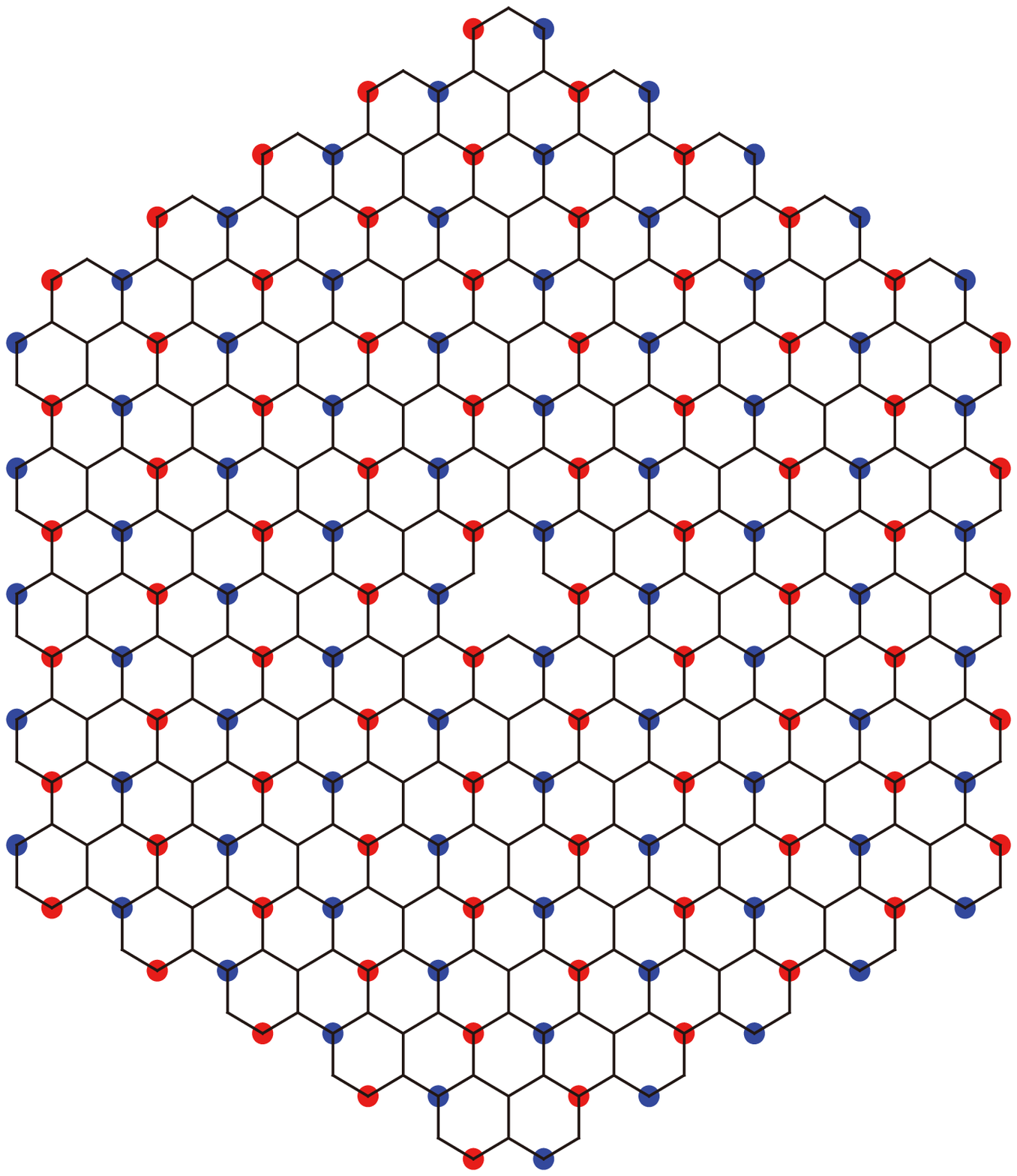}
\end{center} 
\end{minipage}
\end{tabular}
\caption{(Color) Wave function of (a) QLZM and (b) S3ZM in case of $N_{\rm arm}=5$. 
The size of the circles shows the absolute value of the wave function, and the red and blue colors show the sign of the wave function on each site.}
\label{fig:VANG_NA5}
\end{figure}

When we apply TBM, by adopting the numerical diagonalization method, we always see two degenerated zero-energy modes at $E = 0$ (Fig.~\ref{fig:GapDelta}(a)). One is QLZM (Fig.~\ref{fig:VANG_NA5}(a)) and the other is S3ZM (Fig.~\ref{fig:VANG_NA5}(b)). Supposing that the vacancy is on the B-site of the molecule, QLZM has amplitudes only on the A-sites, which is the non-bonding characteristic. There is a finite energy gap $t\Delta$ between the two zero-energy modes and the other modes. This gap $\Delta$ normalized by $t$ is approximately proportional to $1/N_{arm}$. This result suggests that the energy levels around the Fermi energy are determined as a confined state of the Dirac electron in VANG. Thus the size of the gap can be controlled by the choice of the size of the molecule (Fig. \ref{fig:GapDelta}(b)). 

As is well known\cite{Longuet-Higgins1950},  the zero mode wave function is easily determined for TBM on a bipartite graph. 
The TBM wave function is determined by its amplitude $\psi_{i,l}$ 
at an $i$-th site belonging to the $l$-th sublattice 
($l=A$ or $B$), where a partial site index $i$ on each sublattice runs from 
1 to $N_C/2$, and the function may be given in a vector form 
of $\psi_{A}= {}^t(\psi_{1,A},\cdots,\psi_{N_C/2,A})$ and 
$\psi_{B}= {}^t(\psi_{1,B},\cdots,\psi_{N_C/2,B})$. 
The determination equation may be summarized in a matrix form as, 
\begin{equation}
\left( 
\begin{array}{cc} 0 & H_{AB} \\ H_{BA} & 0 \end{array}
\right) 
\left(
\begin{array}{c} \psi_{A} \\ \psi_{B} \end{array}
\right)
= E
\left(
\begin{array}{c} \psi_{A} \\ \psi_{B} \end{array} 
\right), 
\end{equation}
where $H_{AB}$ is a matrix representing electron transfer 
from a B-site to an A-site and $H_{BA}= {}^t H_{AB}$ and 
$E$ is the eigen energy. 
Since the Schr\"{o}dinger equation is decoupled when $E=0$, 
we only need to find non-zero amplitudes on the A-sites, 
and the determination equation becomes $H_{BA}\psi_{A}=0$. 
Since the transfer is assumed to be finite on each pair of 
nearest-neighbor sites, the condition for $\psi_{A}$ is described by 
a zero-sum rule, which tells that 
the summation of the amplitudes on A-sites around a B site has to be 0. 

\begin{figure}[tb]
\begin{tabular}{cc}
\begin{minipage}{0.5\hsize}
\begin{center}
\begin{flushleft}(a)\end{flushleft} 
\includegraphics[width=4cm]{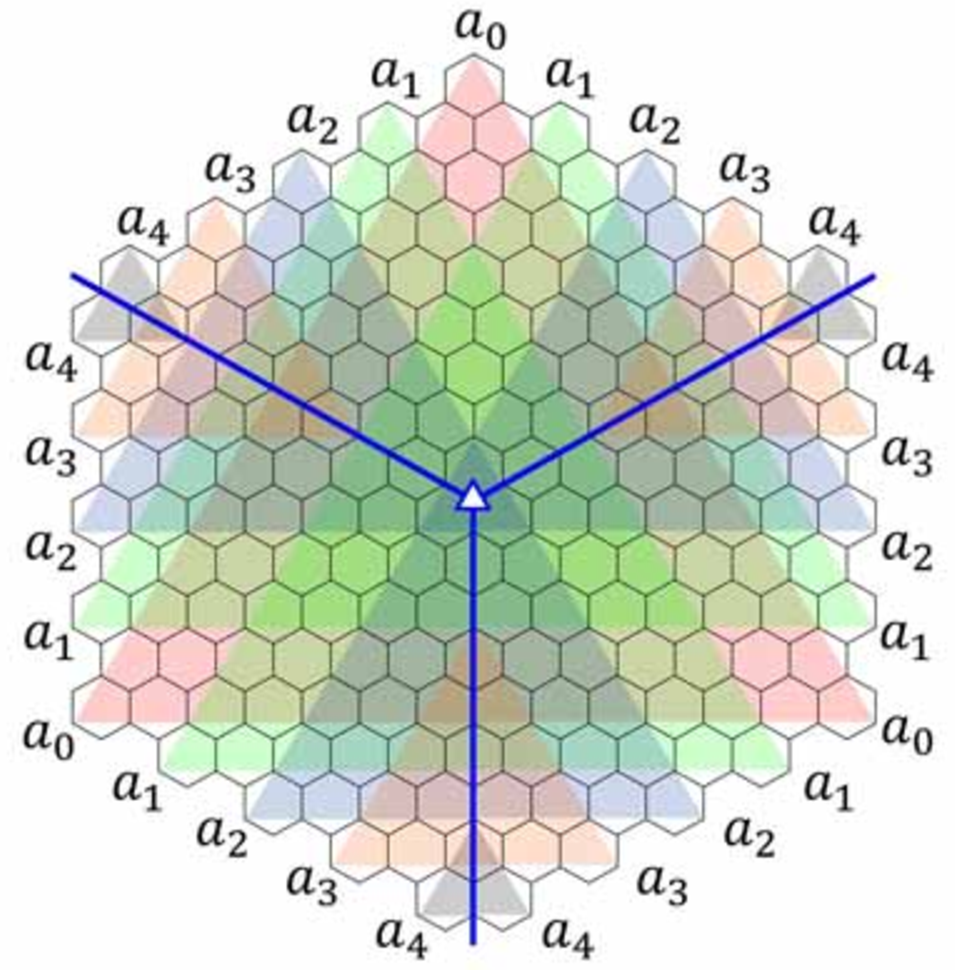}
\end{center}
\end{minipage}
\begin{minipage}{0.5\hsize}
\begin{center}
\begin{flushleft}(b)\end{flushleft}
\includegraphics[width=4cm]{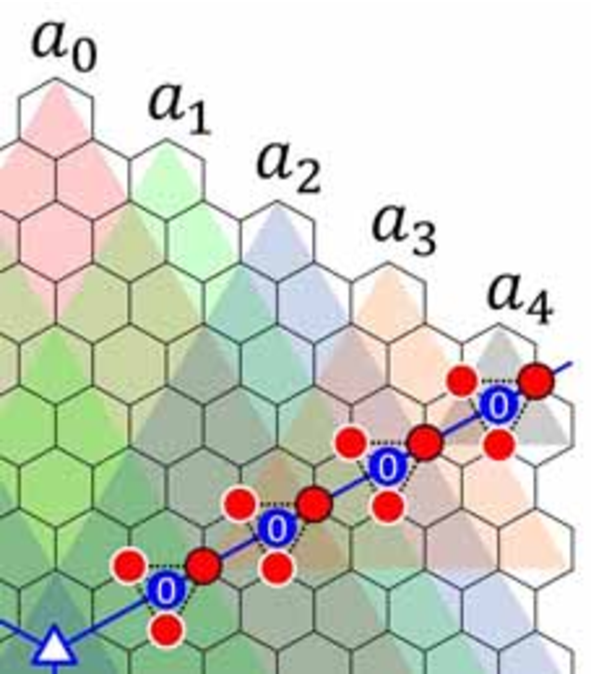}
\end{center} 
\end{minipage}
\end{tabular}
\caption{(Color) (a) Overall section of a sequence of triangles in the case of $N_{\rm arm}=5$. We have five triangles derived from $a_0, ..., a_4$, which touch each other along mirror planes reaching from the center to DZCs (the blue lines). 
(b) By considering the zero-sum rule and the symmetry along a mirror plane, 
a ratio of the values on the A-sites along the mirror plane is determined at each set of three A-sites (red circles) connected by dashed lines. In this example, $4(=5-1)$ equations are derived from the conditions along the mirror plane.}
\label{fig:Condition_Mirror}
\end{figure}

When we consider the $C_{3v}$ symmetry of the $\pi$ network (Fig.~\ref{fig:_C3vSymNA5}) and the above rule for the zero mode wave function, we can determine the value of the wave function of the zero mode on each A-site through the following steps:\\
1. Considering the zero mode on each end of the A-sites along each armchair edge, we write down a sequence forming Pascal's triangle with signs, $(-1)^{n}{}_nC_k$, on A-sites in the bulk of the molecule. Here $n$ is the row of the triangle and ${}_nC_k$ is the binomial coefficient. Specifically, we consider a region bounded by blue lines, which gives a mirror plane, in Fig.~\ref{fig:Condition_Mirror}~(a). Each triangle and any superposition of the triangles then satisfies the 0 sum rule on every B-site in the region. We assign amplitudes to the sequences as $a_0, ...,a_{N_{\rm arm}-1}$ to derive a self-consistent solution.\\
2. We search for a solution for the A$_1$ representation. Considering the consistency condition along the mirror planes (Fig.~\ref{fig:Condition_Mirror}(a)), we obtain simultaneous determination equations for the coefficients. We obtain $N_{\rm arm}-1$ equations for $N_{\rm arm}$ triangles (Fig.~\ref{fig:Condition_Mirror}(b)). Since a system of $(N_{\rm arm}-1)$ homogeneous linear equations for $N_{\rm arm}$ variables has a non-trivial solution, we have a QLZM solution in general. The details of the derivation is found in Appendix~\ref{Determination_equation_QLZM} and \ref{Construction_partial_wave_QLZM}. 

In the case of $N_{\rm arm}=5$, this set of equations is as follows: 
\renewcommand\arraycolsep{1.2pt}
\renewcommand\arraystretch{1.5}
\begin{eqnarray}
\left\{
\begin{array}{cc}
     \,\;140a_0 -120a_1 +27a_2   -\,\;a_3&=0, \\
    \;-15a_0   +\,\;36a_1 -30a_2 +6a_3&=0, \\
    \;\;\;\;\;\;\;\;\;\;\;\;\;\;\;-2a_1  \;+\;9a_2 -9a_3 \; +\,\;a_4&=0,\\
      \;\;\;\;\;\;\;\;\;\;\;\;\;\;\;\;\;\;\;\;\;\;\;\;\;\;\;\;\;\;\;\;\;\;\;\;\;2a_3\;-3a_4&=0.
\end{array}
\right.
\label{eq:sim_eq_5}
\end{eqnarray}
See Appendix~\ref{Determination_equation_QLZM} for a general form of the simultaneous determination equations. 

We can obtain a unique solution to these equations by taking $a_0$ as a normalization constant. Excluding the normalization, the solution can be given by a series of fractions. The zero mode is an A$_1$ representation, which is QLZM with a quasi-localized wave function. For a VANG molecule, we have another solution of S3ZM that is an A$_2$ representation. The solution as exemplified by Fig.~\ref{fig:VANG_NA5}(b) is identified as a wave function of the Dirac point for graphene. Thus, the state is an extending state. An existence proof on S3ZM for arbitrary VANG is given in Appendix~\ref{Determination_S3ZM}. 

\section{Electronic properties of VANG}
Electronic properties coming from the structural symmetry are robust against slight deformation. In reality, V$_{111}$ is known to have a lowered symmetry of $C_{1v}$\cite{Ziatdinov2014}, which was consistent with our DFT-LDA result. Even if so, non-bonding character of the zero modes are protected against the perturbative symmetry reduction from $C_{3v}$ to $C_{1v}$. This is because the molecular levels appear in discrete spectrum and the non-bonding character comes from topological nature of the $\pi$ network. 

The special symmetric properties of VANG result in the appearance of interesting phenomena, especially when we consider the many body effect between electrons in these modes. Because of the quasi-localized nature of QLZM, an electron correlation effect should make the QLZM a singly occupied molecular orbital (SOMO), so that the S3ZM becomes a SOMO as well. The degenerate zero-energy modes cause magnetic activity. We can further analyze the magnetic properties.

Interestingly, for the TBM model of the molecule, supposing that the inter $\pi$-orbital exchange integral $J$ is negligible compared to the on-site (intra single $\pi$ orbital) Coulomb correlation energy $U$, the estimated direct ferromagnetic exchange interaction between the two modes become vanishingly small compared to direct Coulomb integrals. This comes from the symmetry of the molecular orbitals (MOs), where the QLZM has non-zero amplitudes only on A-sites, whereas S3ZM is non-zero only on B-sites. 

The $\pi$ network of VANG is bipartite, and the number of A subgraph sites is the same as that of B subgraph sites ($N_{\rm A}=N_{\rm B}$). 
Applying the Hubbard model as an effective description, the well-known theorem\cite{Lieb1989} determines that the ground state is a singlet. However, because of the degeneracy between the QLZM and the S3ZM, the stability of the singlet ground state appears only in a higher order correction of a smaller order than $O(U^2)$ and a low-energy triplet state exists. In fact, using perturbation theory, we can derive an indirect anti-ferromagnetic exchange interaction between QLZM and S3ZM within the Hubbard model. 

Here we evaluate the low-lying excitation energy using the local spin-density approximation (LSDA). Considering the VANG molecule C$_{60}$H$_{24}$ in Fig.~\ref{fig:DFT_VANG_NA2}, the lowest variational state in LSDA becomes an anti-ferromagnetic (AF) spin state with spin-up density in QLZM 
and spin-down density in S3ZM. Using a constraint on the total spin moment, we estimate the lowest spin triplet state of C$_{60}$H$_{24}$, which has an energy higher by about 70 meV than the AF solution.

We may consider the LSDA solution as a variational estimation of the true system, where quantum mechanical exchange splitting should be the order of the LSDA mean-field splitting. An energy gap $\Delta$ separates these zero modes from the other MOs. Even if we consider magnetic perturbations, the value of $\Delta$ is of the order of $0.1t$, which amounts to $0.3$eV, and is larger than the expected magnetic interaction, which would not exceed a few tens of meV. Therefore, degeneracy in the ground state of VANG would appear as nearly degenerate magnetic multiplets, which allows us to design curious quantum-information-device applications.

\begin{figure}[t]
\begin{center}
\includegraphics[height=8cm]{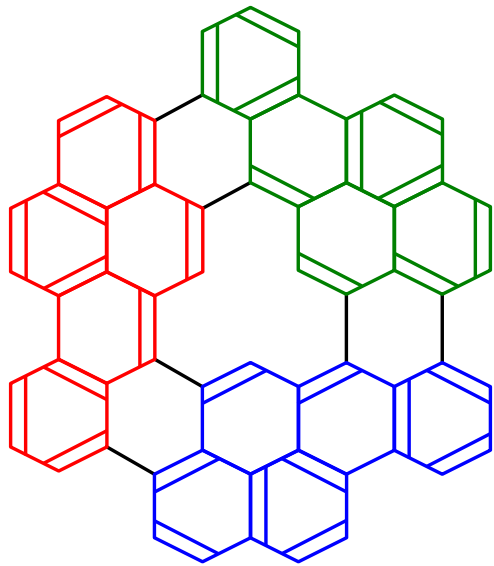}
\end{center}
\caption{(Color-online) Kekul\'{e}'s structure for a VANG molecule C$_{60}$H$_{24}$ of $N_{\rm arm}=2$.}
\label{fig:Kekule_NA2}
\end{figure}

We comment on stability of VANG as a Kekul\'{e} PAH. We can find several Kekul\'{e} structures for each VANG molecule. The concluded singlet ground state is consistent with this observation. Furthermore, a typical Kekul\'{e} structure demonstrates that a VANG molecule may be recognized as a polymer of three monomers: {\it e.g.} three benzo[$a$]pyrenes for C$_{60}$H$_{24}$ (Fig.~\ref{fig:Kekule_NA2}). 
This result strongly suggests that a synthetic route of VANG may be found, although one might design a synthesis path using appropriate derivatives of benzo[$a$]pyrenes. 

Although the structure should have only inactive (or terminated) $\sigma$ bonds at the vacancy of VANG, the vacancy may be a mono-hydrogenated atomic vacancy known as V$_1$. In that structure, two of the dangling bonds make a weak covalent bond, and the remaining carbon $\sigma$ orbital is terminated by a hydrogen atom creating a $\sigma$ bond. When the topology of the $\pi$ network is identified to be equivalent to that in Fig.~\ref{fig:_C3vSymNA5}, we include all of the molecules possessing this $\pi$ topological network in the category of VANG molecules.

\section{Conclusion}
In conclusion, by considering the zero mode written in an algebraic form, we have shown the following. There is a group of molecules (VANG) whose characteristic $\pi$ network should show the following properties. The $\pi$ electron system has (at least nearly) two-fold degenerate zero-energy modes in the highest occupied state. One is the vacancy-centered zero mode and the other is the $\sqrt{3} \times \sqrt{3}$ mode. Each zero mode should be singly occupied by an electron owing to the electron correlation effect. The stability of the atomic structure as well as the unique electronic structure should be protected against weak external perturbations. 

\section*{Acknowledgment}
The authors greatly thank stimulating discussions and fruitful comments by Professor I. Maruyama, Professor T. Enoki, Professor K. Takai. They would like to thank Enago for the English language review. The calculations were partly performed in the computer centers of Research Center for Information Technology, Kyushu University and Institute for Solid State Physics, University of Tokyo. This work is supported by the Grant-in-Aid for scientific research (No. 23540408, No. 26107526, and No. 26400357) from MEXT, Japan.  

\appendix\label{appendix}
\section{The simultaneous determination equations of QLZM}
\label{Determination_equation_QLZM}

In this appendix, let $N=N_{\rm arm}$. We have variables $a_n$ with 
$n=0,1,\cdots, N-1$. We have $N-1$ linear equations of $a_n$. 
The determination equations of $a_n$ are given in the next form. 
\begin{equation}
\sum_{n=0}^{N-1} C_{m,n} a_n = 0.
\label{Determination_N_VANG}
\end{equation}
Here $m$ runs from 1 to $N-1$. 
To determine $n$-th coefficient $C_{m,n}$ for the $m$-th equation, 
with two integers $l$ and $k$, we define an expression $BC(l,k)$. 
\begin{eqnarray}
BC(l,k) = \left\{
\begin{array}{ll}
\frac{l!}{k!(l-k)!}  & (l > k > 0) \\
1 & (k = l \lor k = 0)  \\
0 & (k > l \lor 0 > k)
\end{array}
\right.
\end{eqnarray}
The expression of $C_{m,n}$ is given in the next forms. \\
\hspace*{5mm} For $n\neq0$ 
\begin{eqnarray}
C_{m,n} &=& (-1)^{- 1 - n - m } \nonumber \\
&\times&[2BC(2N - 1 - n - m, N - 1 - 2n + m) \nonumber \\
&&+ BC(2N - 1 - n - m, N - 2n  + m) \nonumber \\
&&+ 2BC(2N - 1 - n - m, N - 1 + n + m) \nonumber \\
&&+ BC(2N - 1 - n - m, N + n  + m)],
\end{eqnarray}
\hspace*{5mm} for $n=0$
\begin{eqnarray}
C_{m,n} &=& (-1)^{- 1 - n - m } \nonumber \\
&\times&[2BC(2N - 1 - n - m, N - 1 - 2n + m) \nonumber \\
&&+ BC(2N - 1 - n - m, N - 2n  + m) 
].
\end{eqnarray}
Some solutions are listed in Table~\ref{tab:table1}. 

\begin{table}[h]
\caption{\label{tab:table1}
Ratio of the coefficients of each triangle $a_1/a_0$, $a_2/a_0$, $\cdots$, 
$a_{N-1}/a_0$. Every value is always positive, and $a_{N-2}/a_{N-1} = 3/2$ 
due to the conditions around the vertex of the hexagonal structure.
}

\begin{tabular}{c|ccccc}
  
\hline
$N_{\rm arm}$  & 2 & 3 & 4 & 5 & 6 \\ 
\hline\hline


$a_1/a_0$ &   2/3 &  27/25& 432/319& 3025/2001& 1830114/1148311\\

$a_2/a_0$ &          & 18/25 & 390/319& 4225/2668&  2086903/1148311\\

$a_3/a_0$ &          &          & 260/319 &  3595/2668& 5990852/3444933\\

$a_4/a_0$ &         &           &             & 3595/4002 & 1655850/1148311\\

$a_5/a_0$ &         &           &             &               & 1103900/1148311 \\
\hline
\end{tabular}
\end{table}

\section{Construction of QLZM wave function}
\label{Construction_partial_wave_QLZM}
As shown in Fig.~\ref{fig:Condition_Mirror}, we have three bulk regions 
surrounded by the blue half lines representing positions of mirror planes. 
When the zero-sum rule is satisfied at each B-site in one of the bulk regions, 
the bulk zero-sum rule, which will be defined later, is satisfied. 
This condition is easily met 
by the form of Pascal's triangle with the signs, $(-1)^{n} {}_nC_k$ on 
A-sites within the region. Note that the construction allows us 
to determine amplitudes on A-sites along the blue lines. 

As explained in the main text, we have a kind of partial waves formed by 
the signed Pascal triangle structure. 
The function satisfies the bulk zero-sum rule. 
Furthermore, we may assign symmetry for each partial wave. 
The remaining problem is to satisfy the zero-sum rule on 
B-sites along the blue lines. 
This is summarized in the self-consistent equations, whose general 
form is given by Eq.~\ref{Determination_N_VANG} with definitions 
in Appendix~\ref{Determination_equation_QLZM}. 

When we determine relative amplitudes following the determination 
equations in Eq.~\ref{Determination_N_VANG}, we have 
the total wave function of QLZM satisfying the zero-sum rule globally 
in the VANG structure (the global zero-sum rule). 

\begin{figure}[tb]
\begin{tabular}{cc}
\begin{minipage}{40mm}
\begin{center}
\includegraphics[height=4cm]{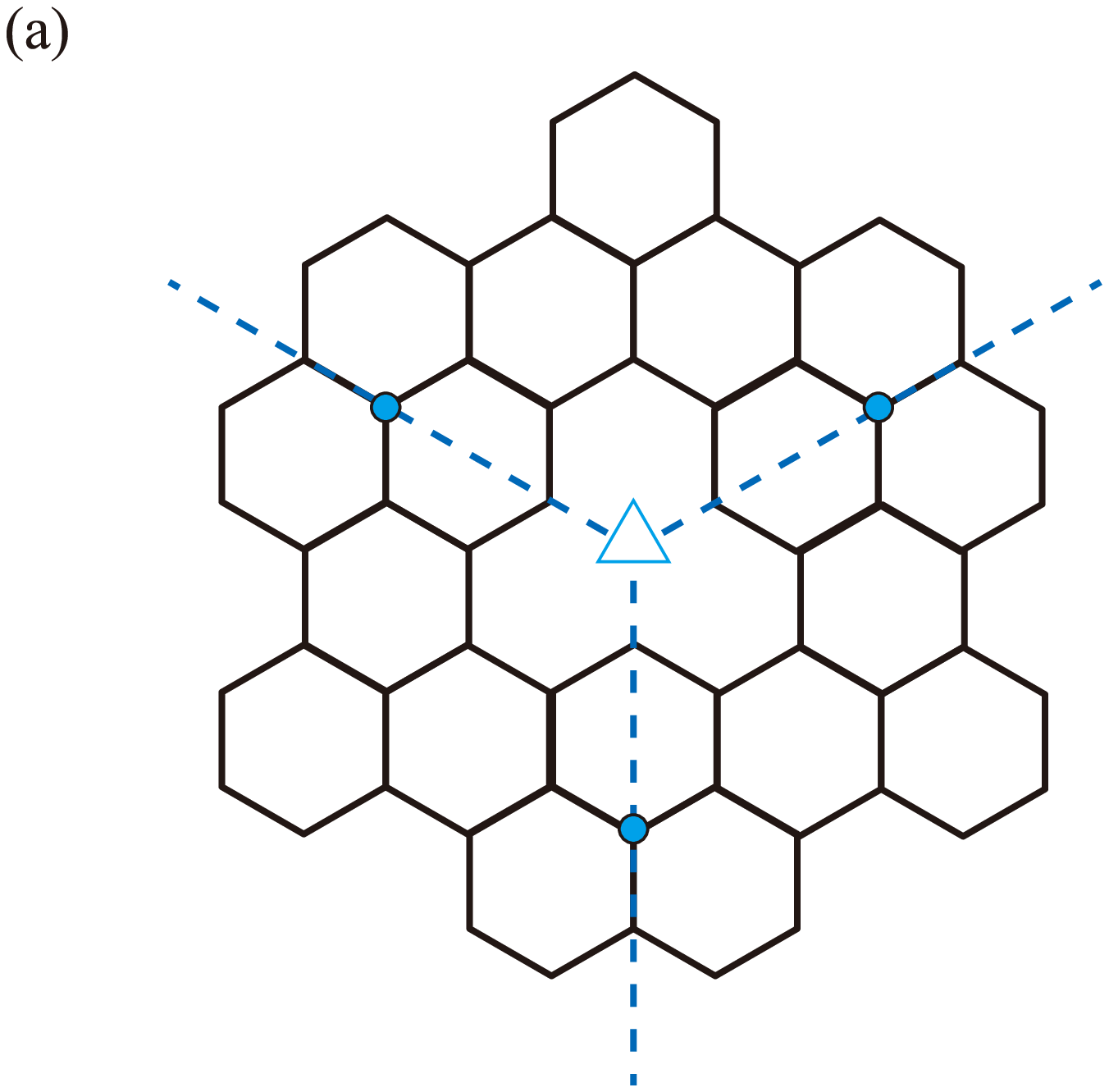}
\end{center}
\end{minipage}
\begin{minipage}{40mm}
\begin{center}
\includegraphics[height=4cm]{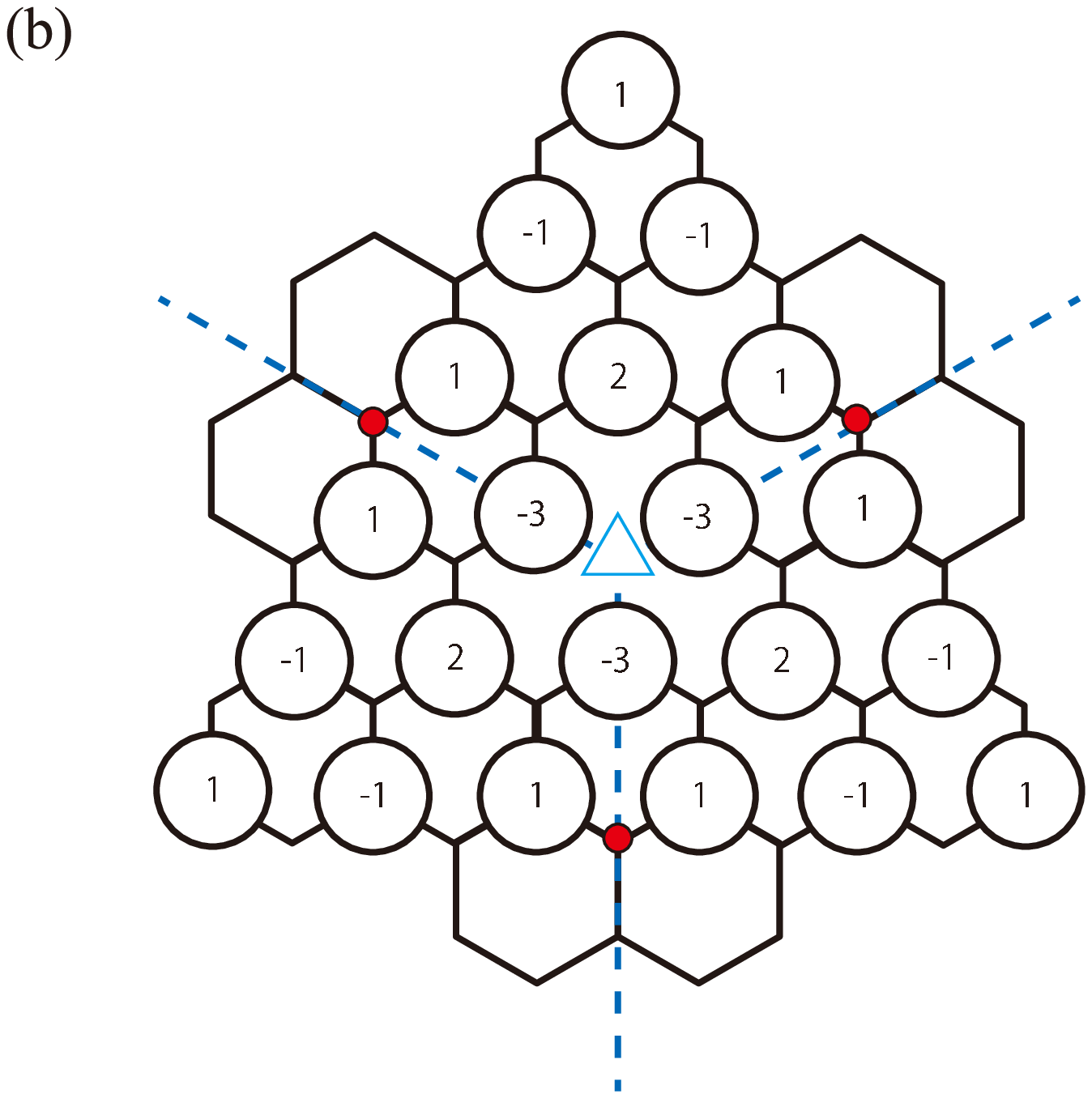}
\end{center} 
\end{minipage} \\ \\
\begin{minipage}{40mm}
\begin{center}
\includegraphics[height=4cm]{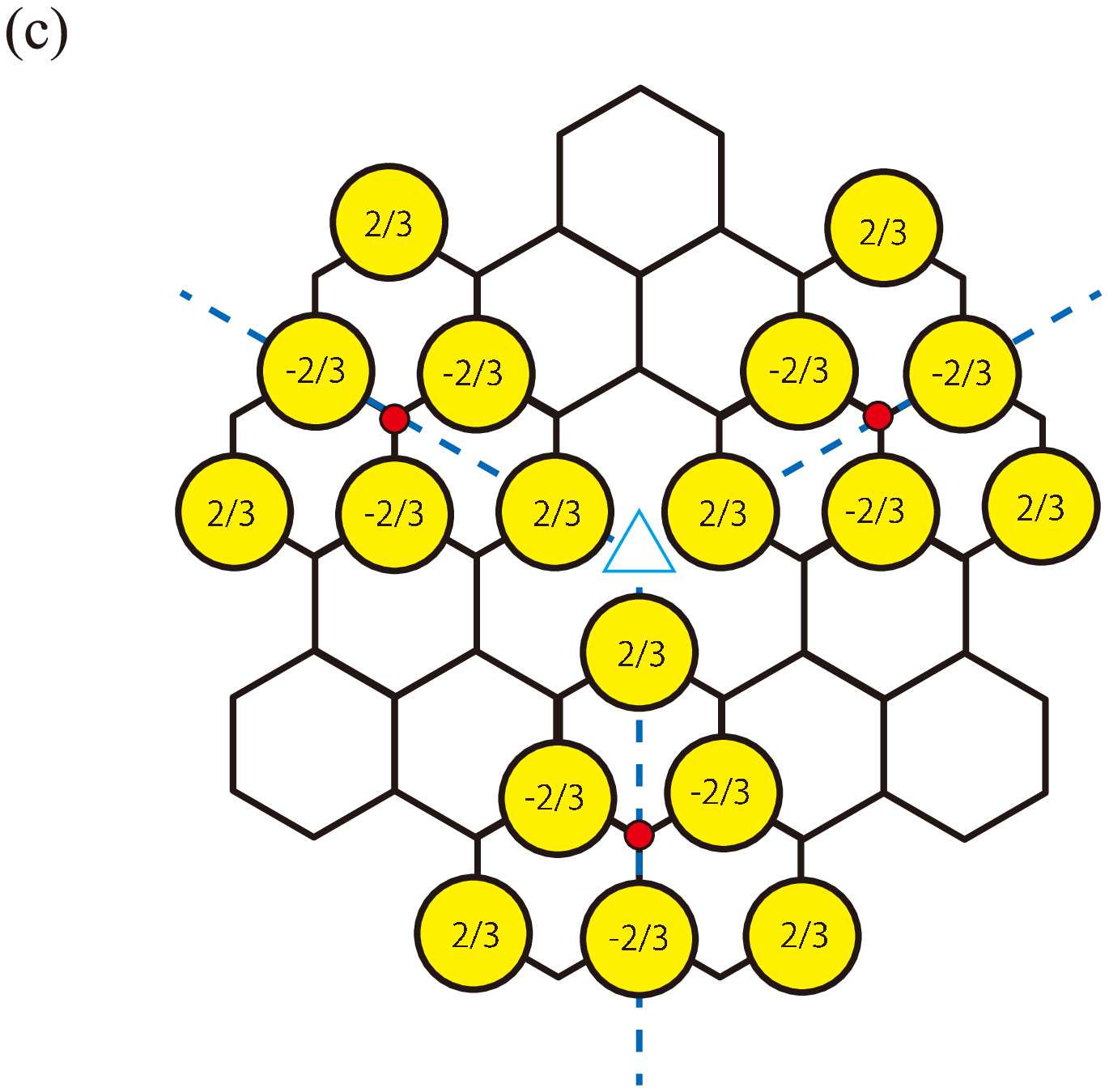}
\end{center}
\end{minipage}
\begin{minipage}{40mm}
\begin{center}
\includegraphics[height=4cm]{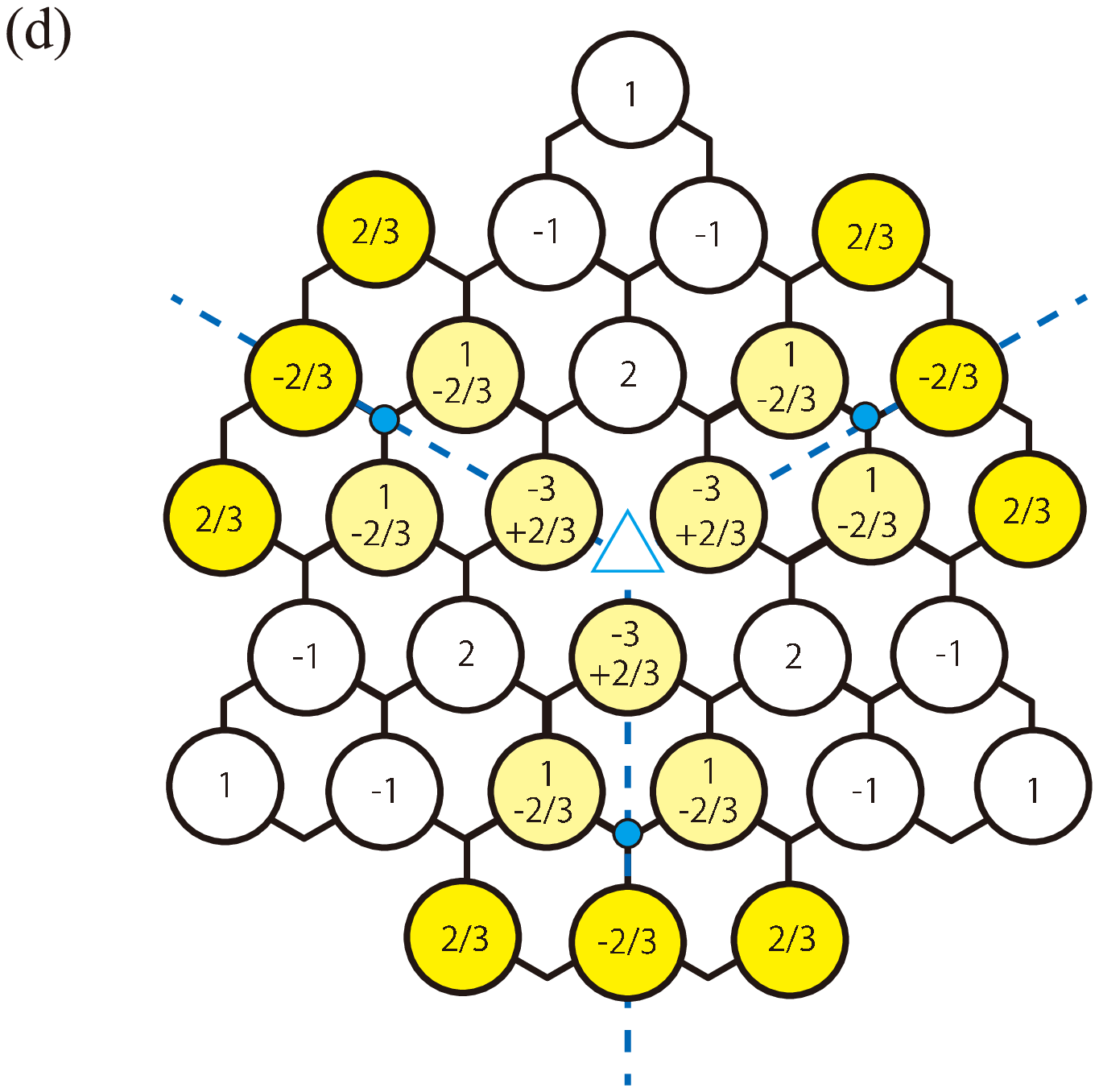}
\end{center} 
\end{minipage} \\ \\
\begin{minipage}{40mm}
\begin{center}
\includegraphics[height=4cm]{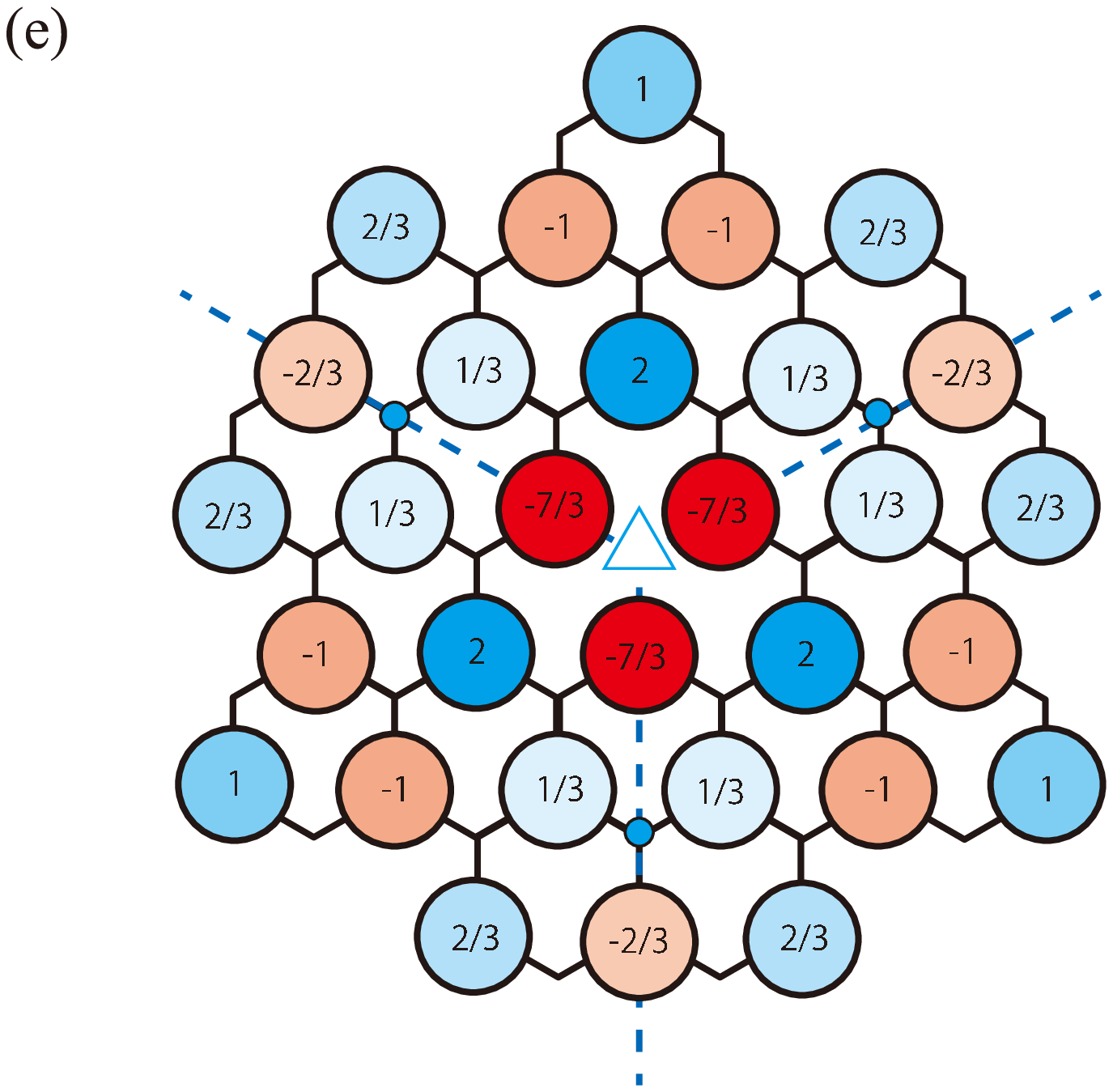}
\end{center}
\end{minipage} \\ \\
\end{tabular}
\caption{(Color) 
Construction of the QLZM wave function 
for TBM of a VANG molecular structure with $N_{\rm arm}=2$. 
(a) At the center denoted by the blue triangle, 
there is a vacancy site without any $\pi$ orbital. 
Blue dashed lines represent the mirror planes. 
In each region surrounded by two of blue half lines, 
every B-site is considered as a site in the bulk region. 
B-sites on the blue lines correspond to the sites 
colored blue in Fig.~\ref{fig:Condition_Mirror}. 
(b) A part of the wave function (a partial wave function) 
derived by signed Pascal's triangles. 
The form is derived to satisfy the bulk zero-sum rule explained in the text. 
The determined amplitudes do not satisfy the zero-sum rule on the B-sites along 
the blue lines, which is colored red. Note that the center of the vacancy 
site is just a void without a B-site, so that no additional constraint appears 
by the void site. 
(c) Another partial wave function, whose amplitude has been 
adjusted by the secular equations explained in the text. 
By summing these two sets of amplitudes given in (b) and (c), we have 
a full wave function represented in (d), where 
the zero-sum rule is satisfied also at each blue B-site on the blue lines. 
(e) The determined total wave function for $N_{\rm arm}=2$. 
The topology of amplitudes with respect to its sign and 
relative strength is the same as QLZM of Fig.~\ref{fig:DFT_VANG_NA2} (a). 
}
\label{fig:WF_NA2}
\end{figure}

We show the wave function of QLZM determined by 
the single-orbital tight-binding model for VANG with $N_{\rm arm}=2$ 
in Fig.~\ref{fig:WF_NA2}. By adopting the determined relative amplitudes 
of $a_1/a_0=2/3$, we can actually determine the non-bonding wave function. 

Thus, construction of the QLZM wave function requires us to derive 
1) partial waves to expand the QLZM wave functions 
and 2) a proof of self-consistent conditions summarized 
as Eq.~\ref{Determination_N_VANG} as a result of 
the zero-sum rule at every B-site. 
Now, we describe details of the construction of the partial wave. 

\begin{figure}[tb]
\begin{tabular}{cc}
\begin{minipage}{40mm}
\begin{center}
\includegraphics[height=4.cm]{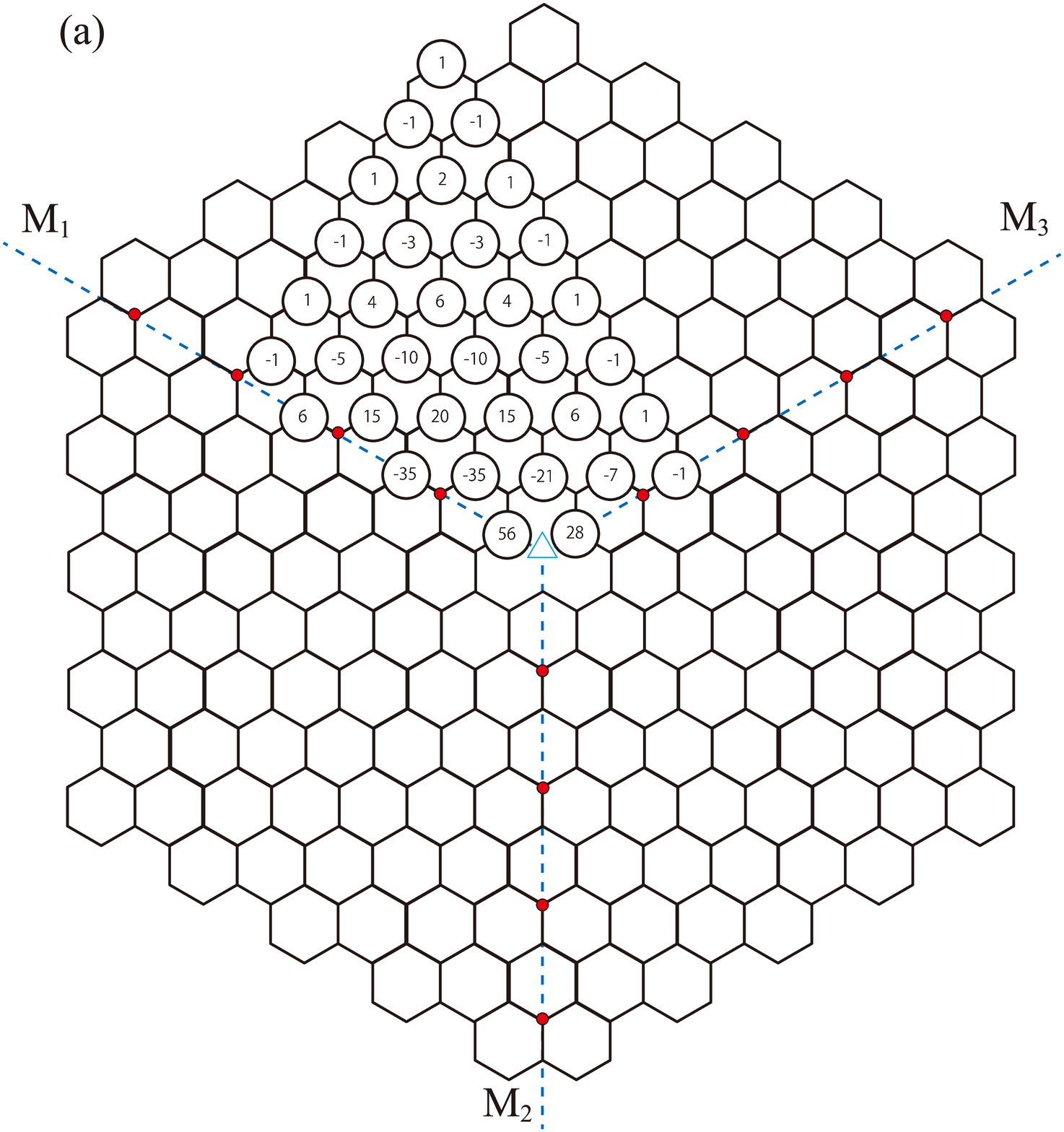}
\end{center}
\end{minipage}
\label{fig:WF_NA5_a}
\begin{minipage}{40mm}
\begin{center}
\includegraphics[height=4.cm]{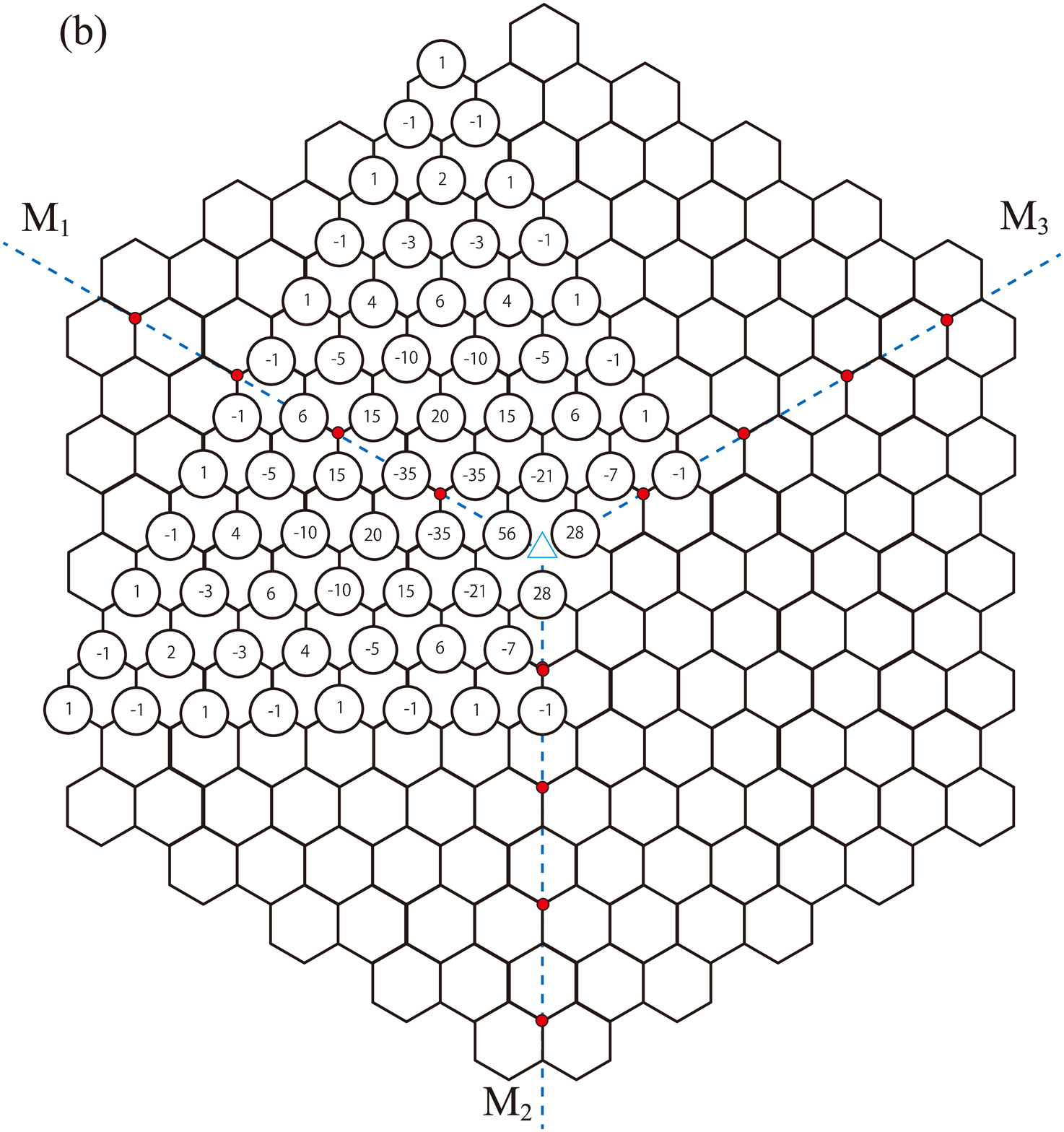}
\end{center} 
\end{minipage}
\label{fig:WF_NA5_b}
 \\ \\
\begin{minipage}{40mm}
\begin{center}
\includegraphics[height=4.cm]{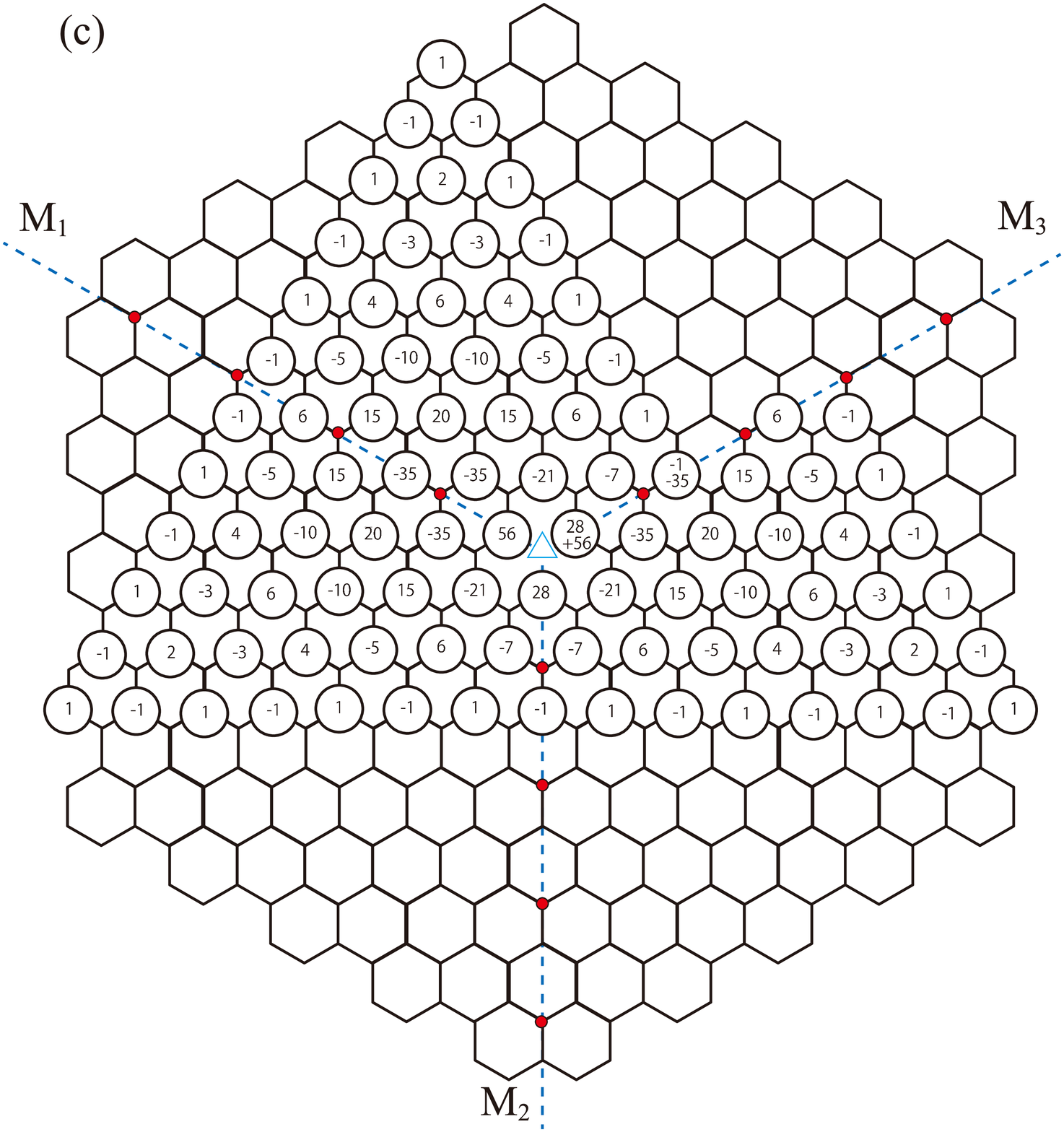}
\end{center}
\end{minipage}
\label{fig:WF_NA5_c}
\begin{minipage}{40mm}
\begin{center}
\includegraphics[height=4.cm]{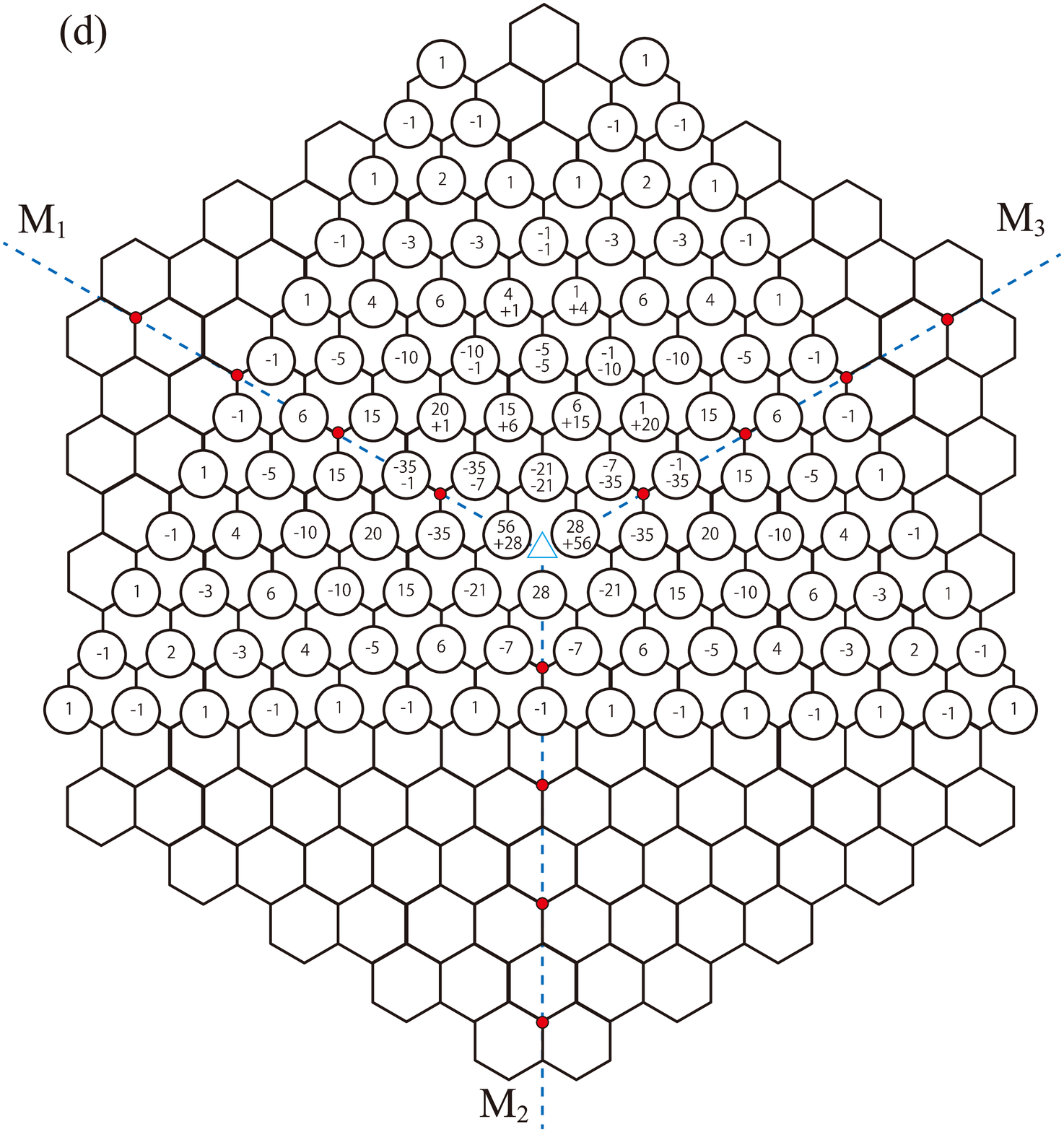}
\end{center}
\end{minipage}
\label{fig:WF_NA5_c}
 \\ \\
\begin{minipage}{40mm}
\begin{center}
\includegraphics[height=4.cm]{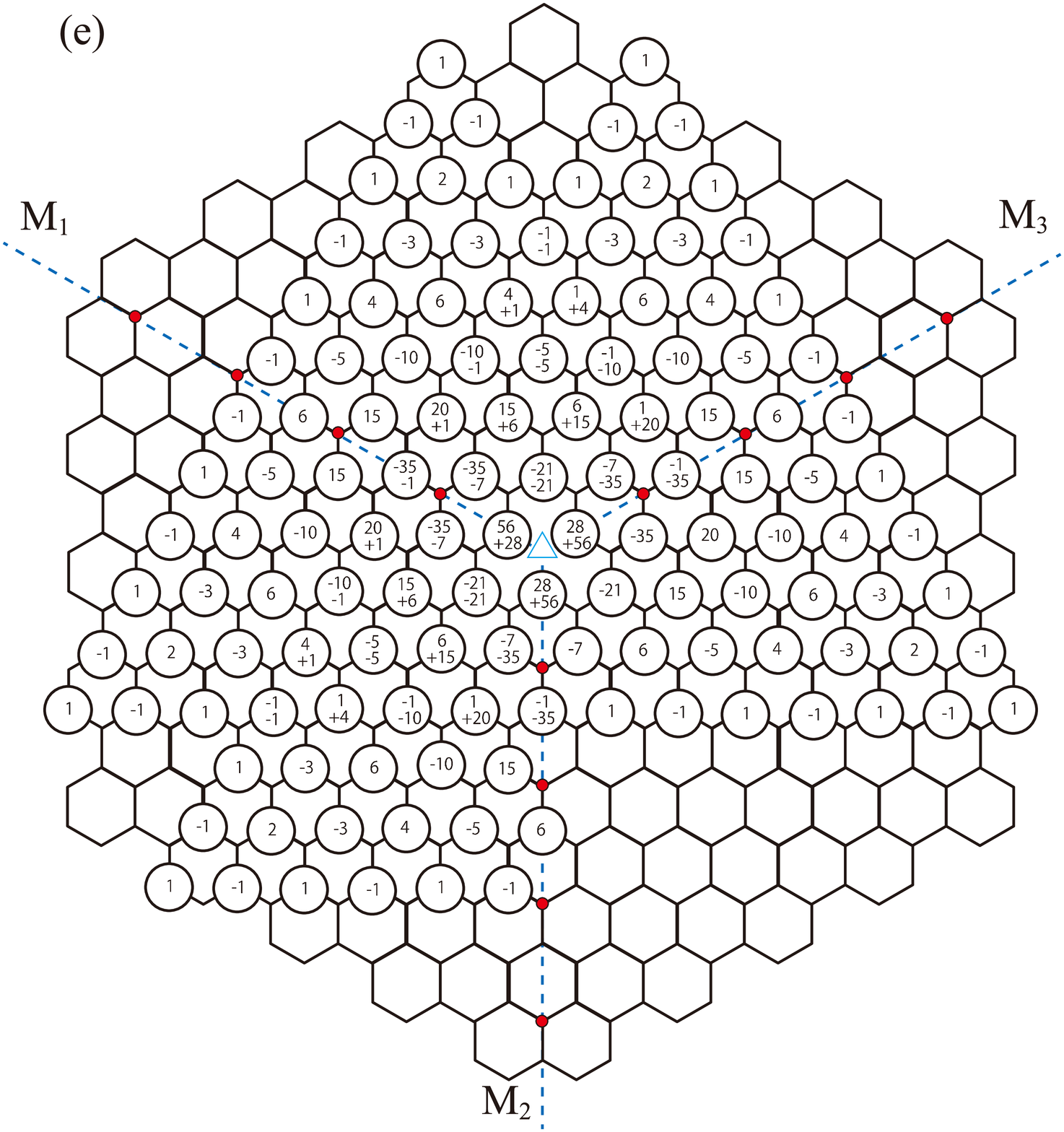}
\end{center}
\end{minipage}
\label{fig:WF_NA5_c}
\begin{minipage}{40mm}
\begin{center}
\includegraphics[height=4.cm]{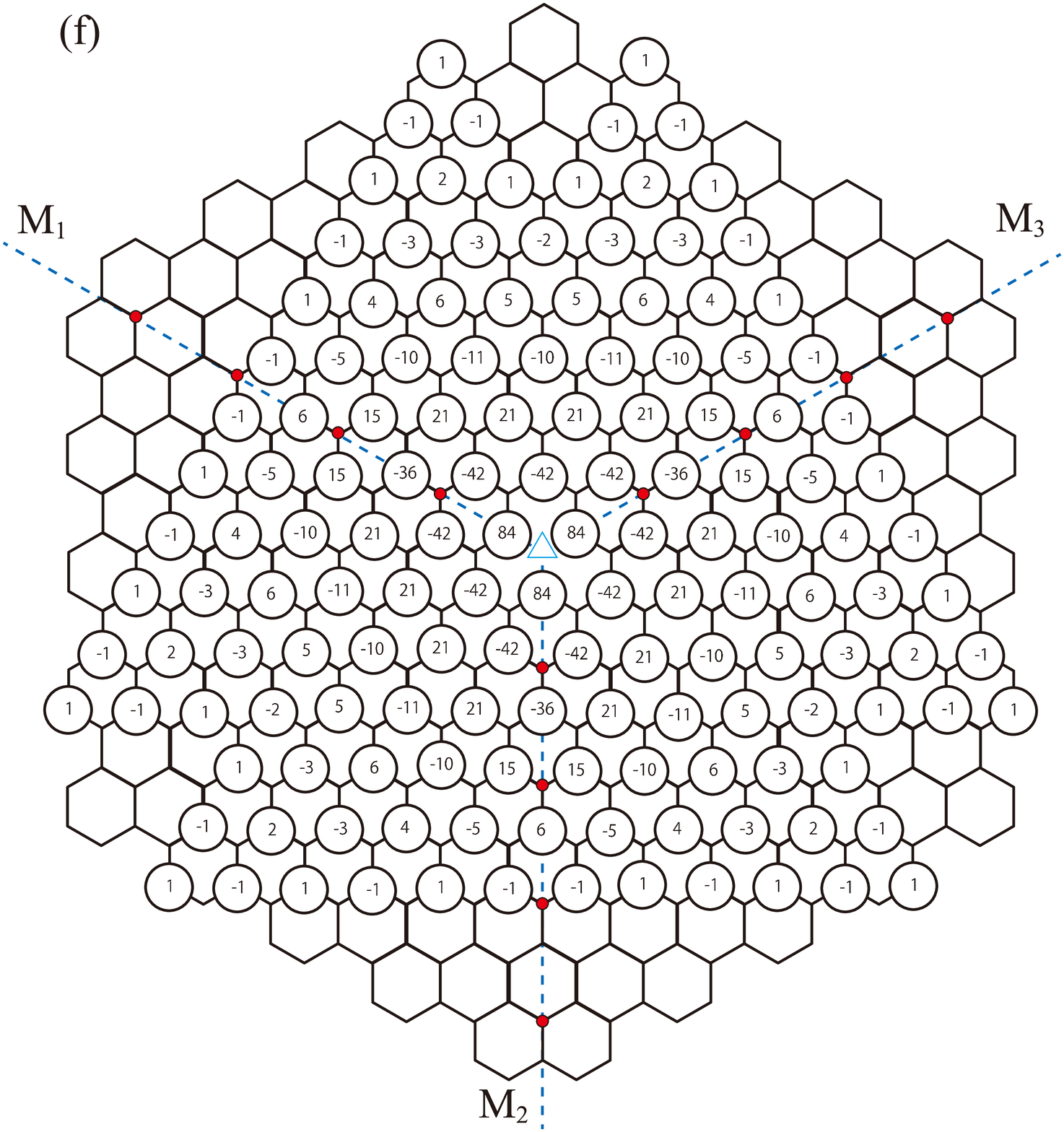}
\end{center} 
\end{minipage}
 \\ \\
\end{tabular}
\caption{(Color) Construction of a partial wave function 
of VANG. The example is given by the case of $N_{\rm arm}=5$, 
although the construction method is applicable for any finite $N_{\rm arm}$. 
(a) In a rhomboidal region surrounded 
by two of mirror planes (M$_1$ and M$_3$) whose positions are 
given by blue dashed half lines, 
we write a signed Pascal triangle whose first row is located 
at an edge A-site of VANG. The structure is determined 
so that the zero-sum rule holds in all B-sites inside the rhomboid region. 
Here, we do not consider the zero-sum rule at each boundary B-site 
on the blue lines, which are represented by red circles. 
(b) The wave is extended to a nearest neighbor rhomboid region, 
where the signed Pascal triangle structure is found to be a mirror 
image of the original. Except for the boundary B-sites (red circles), 
the zero-sum rule holds within the second rhomboid. 
(c) The extension is proceeded to the third rhomboid, where new 
amplitudes on the boundary A-sites on a blue line (M$_3$) 
are added to the original in order 
to satisfy the zero-sum rule in the third rhomboid. 
(d) When the extension is further considered in the first rhomboid region, 
we have another triangle for this example. The new amplitudes 
are added to the original. Here, the newly added part is a mirror image 
given by subsequent mirror reflection operations M$_1$, M$_2$, and M$_3$, 
which is not identical to the original except for a triangle starting from 
a single zigzag corner. 
(e) We can proceed the construction using the new amplitude 
determined in the process (d). 
(f) Finally, the amplitudes are given in the whole VANG structure. 
The obtained function satisfying the symmetry of VANG 
is used as a partial wave function. 
The wave is symbolically represented by 
a light green structure in Fig.~\ref{fig:Condition_Mirror}.  
}
\label{fig:WF_NA5_d}
\end{figure}

We demonstrate the method in Fig.~\ref{fig:WF_NA5_d} by a graphical manner 
as well as a general recipe for the construction using the symmetry of VANG below. 
Let us consider a general VANG structure with $N_{\rm arm}\ge 2$. 
Owing to the structure of VANG, we have three mirror planes 
$M_1$, $M_2$, and $M_3$. On each plane, we can draw a blue 
half line starting from the vacancy site to a double zigzag corner (DZC). 
These three blue half lines determine three rhomboid regions on VANG. 
Two edges of the rhomboid are along the arm-chair edges of VANG. 

There are $N_{\rm arm}$ A-sites along one arm-chair edge of VANG. 
Choose an edge A-site along one arm-chair edge boundary, 
where the boundary line intersects a selected mirror plane $M_1$. 
Inside a rhomboid region including the edge and the selected site, 
we draw a signed Pascal triangle, 
where each of two edges of the triangle 
is perpendicular to  one of the blue lines. 

As for the signed Pascal triangle, it actually satisfies 
the zero-sum rule locally. 
The signed amplitudes, which are also given at A-sites on the mirror planes, 
satisfy the zero-sum rule applied at each B-site in the rhomboid region. 
Therefore, the wave of the triangle satisfies the bulk zero-sum rule within a 
rhomboid region. 

Here, for explanation, we define two terminologies, {\it i.e.} 
a bulk zero-sum rule and a global zero-sum rule. 
When a wave satisfies each condition of the zero-sum rule 
at every B-site inside a rhomboid region, 
the wave satisfies the bulk zero-sum rule. 
Along the blue line, we have $N_{\rm arm}-1$ B-sites. 
At these boundary B-sites, we do not impose 
the zero sum rule on the partial wave. 
After construction of partial waves, considering a linear combination 
of the partial waves, we impose the zero sum rule at every B-sites of VANG, 
which is called the global zero sum rule, on the total wave function of QLZM. 

To create the whole structure of the partial wave, 
we may consider mirror reflected amplitudes in the other region. 
Actually, an extension is possible by operating $M_1$ to the triangle 
to obtain a mirror image. Except for the boundary B-sites along 
blue lines, the extended wave satisfies the bulk zero-sum rule 
in each rhomboid region. 

Since we are aiming at construction of an $A_1$ representation, 
we proceed to extend the partial wave construction 
by applying $M_1$, $M_2$, and $M_3$, successively. 
For any structure on a finite system, 
the operation of the point group forms a finite representation 
of a cyclic group. In $C_{3v}$, we may have maximally 
six elements produced by $M_1$, $M_2$, $M_3$, 
$M_1$, $M_2$, and $M_3$ in this order. 
Our construction requires the zero-sum rule to be satisfied 
except for the red-colored B-sites at the boundary, 
so that the produced wave is connected in that sense. 
Interestingly, after the operation of $R=M_3M_2M_1$, 
except when the triangle is written from 
the single-zigzag-corner (SZC) site, 
the product does not match with the original triangle in general. 
Only when $R^2$ is applied, the triangle is reproduced. 
In that sense, we have connectivity around the vacancy site, 
which is an origin of a non-trivial topological 
nature of QLZM discussed later. 

When we have an unmatched triangle by $R$ operation, 
the amplitudes of two Pascal triangles are summed in each region. 
As a result, we have a partial wave for each A-site along 
the arm-chair edge. 

Since there are $N_{\rm arm}$ A-sites along one edge of the rhomboid region, 
because Pascal's triangle having a first row at one of $N_{\rm arm}$ A-site 
is independent from the other triangle which does not have amplitude 
at that site, there are $N_{\rm arm}$ partial waves for each VANG, 
where each of the wave thus constructed is given 
as an A$_1$ representation of the $C_{3v}$ group. 

We can assign an amplitude $a_n$ 
($n=0,\cdots,N_{\rm arm}-1$) for each 
partial wave to write down the global QLZM wave function in any $N_{\rm arm}$. 
A condition for the set of $a_n$ is a zero-sum rule at a B-site along the blue lines. 
Owing to the A$_1$ symmetry for every partial wave, 
each independent self-consistent condition is given by the zero-sum rule 
at a B-site among $N_{\rm arm}-1$ B-sites along one of the blue half line, 
which is expressed by Eq.~\ref{Determination_N_VANG}. 

The connectivity of the partial wave allows us to 
introduce hidden topological nature of QLZM. 
As discussed above, each partial wave has a connectivity 
condition around the center vacancy. 
We may introduce a $2n\pi$ connectivity for the two-dimensional structure 
with the center vacancy. 
Namely, when a wave satisfies a Schr\"{o}dinger-type 
wave equation locally, and if the wave satisfies 
a boundary condition around a vacancy, 
the global connectivity is well defined. 
The total wave may match the consistency 
after the connection condition is followed $n$ terns 
around the vacancy. In that case, we may say that 
$2n\pi$ connectivity is found. 
When $n=1$ connectivity appears, the wave function 
shows trivial nature found {\it e.g.} in ordinary angular momentum 
eigen functions with the spherical harmonics. 
When a higher order connectivity is found as exemplified 
by quantum Hall states or anyon wave functions, non-trivial 
topological nature is concluded. 

In our system, the determination equation is given in 
a discretized matrix equation, and it is simply described 
as the zero-sum rule for the zero energy mode. 
The local zero-sum rule defines the local connectivity of 
the wave. At the red-colored B-sites, we do not impose 
the zero-sum rule on a partial wave. Therefore, we may say that 
the partial wave follows another Schr\"{o}dinger equation 
for a molecular structure with vacancy sites at these red-colored B-sites. 
In the system, since the number of B-sites are reduced from 
the original VANG structure, 
owing to the $|N_A-N_B|>0$ condition, there appear large numbers 
of zero modes as eigen states. The partial waves discussed in 
this appendix correspond to the eigen modes of the structure 
with $|N_A-N_B|>0$. We can say that, using these eigen modes
for the other problem, QLZM of VANG can be expanded in a series. 

As checked by the above discussion, 
the partial wave has $4\pi$ connectivity, 
except when the triangle starts from SZC. 
Since the connectivity with $n>1$ is non-trivial, 
we conclude that QLZM around the vacancy 
of VANG possesses non-trivial topological quantum number of $n=2$. 
The hidden nature may appear in a response to 
a gauge transformation 
induced by a magnetic flux line penetrating the vacancy site. 

\section{S3ZM for a VANG structure}
\label{Determination_S3ZM}
The wave functions of graphene at the Dirac point 
are known to be four-fold degenerate 
with respect to the pseudo-spin index {\it i.e.} the valley index 
represented by K and K' points in the 1st Brillouin zone, 
and the chirality {\it i.e.} A and B sublattices. 
When we consider S3ZM, we need to write 
a chiral wave function whose amplitudes are non-zero 
only on B-sites. There remains pseudo-spin degrees of freedom, 
which are represented by the wave function at the Dirac points 
$\varphi_{K,B}({\bf r}_i)$  and $\varphi_{K',B}({\bf r}_i)$ 
with the position vector ${\bf r}_i$ of each carbon atom 
($i=1, \cdots, N_C/2$). 
We introduce the Cartesian coordinate whose 
origin is at one of B-sites. 
Then, two of basic vectors, ${\bf a}=(a,0)$ and 
${\bf b}=(-a/2,\sqrt{3}a/2)$ with a bond length $a$ may be 
introduced to represent ${\bf r}_i$. 
With these vectors, the k-vectors for two K points 
are ${\bf k}_K = (k_0,\sqrt{3}k_0)$ 
and ${\bf k}_{K'} = (2k_0,0)$ with $k_0=2\pi/(3a)$. 

Although $\varphi_{K,B}({\bf r}_i)$ and 
$\varphi_{K',B}({\bf r}_i)$ are complex valued, 
$\tilde{\varphi}_{1,B}({\bf r}_i)\equiv 
\left\{\varphi_{K,B}({\bf r}_i)
+\varphi_{K,B}({\bf r}_i)\right\}/\sqrt{2}$ and 
$\tilde{\varphi}_{2,B}({\bf r}_i)\equiv 
\left\{\varphi_{K,B}({\bf r}_i)
-\varphi_{K,B}({\bf r}_i)\right\}/\sqrt{2}$ are real valued functions. 
Interestingly, amplitudes of $\tilde{\varphi}_{2,B}({\bf r}_i)$ 
are identical to S3ZM in their distribution. 
Therefore, a remaining point to be shown is that 
the boundary condition of VANG is consistent with 
this zero-energy wave function. 

The amplitude of $\tilde{\varphi}_{2,B}({\bf r}_i)$ is zero at the origin, 
and the zero-sum rule for this wave function is assigned at each A-site, 
so that creation of the center B-site vacancy has no effect 
on the zero-sum rule for this wave function. 
We may cut a graphene structure in a fragment 
without harming the zero-sum rule for 
$\tilde{\varphi}_{2,B}({\bf r}_i)$ by the next manner. 
When one create boundary by cutting C-C bonds, 
a boundary site with two-fold bond connection appears. We should 
consider these boundary sites only. 
If each A-site at the boundary is surrounded by two B-sites 
with finite amplitudes of $\tilde{\varphi}_{2,B}({\bf r}_i)$, 
since the phase of the wave function at one of sites 
has to be opposite to that at the other site, 
the zero-sum rules are not affected. 
Actually, the boundary of VANG defined in the text 
satisfies this rule in general. 
Thus S3ZM appears as an eigen state in every VANG structure. 
In other words, the definition of the VANG structure is 
given to have one QLZM and one S3ZM as eigenstates. 



\end{document}